\documentclass[prl,twocolumn,showpacs,amsmath,amssymb,superscriptaddress]{revtex4}
\usepackage{graphicx}
\begin{document}
\title{Resonant Spin and Charge Hall Effects in 2D Electron Gas
with Unequal Rashba and Dresselhaus Spin-Orbit Couplings under a Perpendicular Magnetic Field}
\author{Degang Zhang}
\affiliation{College of Physics and Electronic Engineering, Sichuan Normal University,
Chengdu 610101, China\\
and Institute of Solid State Physics, Sichuan Normal University,
Chengdu 610101, China}
\affiliation{Texas Center for Superconductivity and Department
of Physics, University of Houston, Houston, TX 77204, USA}
\author{C. S. Ting}
\affiliation{Texas Center for Superconductivity and Department
of Physics, University of Houston, Houston, TX 77204, USA}

\begin{abstract}

We have investigated the complex two-dimensional electron system with unequal Rashba and Dresselhaus spin-orbit interactions in the presence of a perpendicular magnetic field. The spin polarizations are obtained in a wide range of magnetic fields. It is shown that such a system is hard to be magnetized. We also find that the resonant charge and spin Hall conductances occurs simultaneously at a certain magnetic field, at which two (nearly) degenerate Landau levels are filled partly. The resonant Hall effects are universal in this type of semiconductor materials, and could have potential application for semiconductor spintronics.

\end{abstract}


\maketitle

The spin Hall effect in two-dimensional electron gases (2DEG) has been the focus of theoretical and experimental investigations in the condensed matter community due to its potential application in designing quantum devices [1-6]. Such an effect is usually generated by the Rashba or Dresselhaus spin-orbit coupling in semiconductor materials [7,8,9]. In Ref. [10], Shen {\it et al.} discovered an interesting resonant spin Hall conductance in 2DEG with the Rashba spin-orbit coupling in the presence of a perpendicular magnetic field due to the crossing of two nearest neighbor Landau levels. The resonant phenomenon could be observed only if this degeneracy happens at the Fermi level. We note there are a lot of semiconductor materials possessing simultaneously the Rashba and Dresselhaus spin-orbit interactions, which originate from the lack of structure and bulk inversion symmetries [7,8], respectively.
Therefore, the competition among the Rashba and Dresselhaus spin-orbit couplings and the Zeeman energy leads to a complex energy spectrum [11], which could produce novel physical properties in such the materials. The  eigenfunctions associated with the Landau levels are infinite series, distinguishing from those with finite terms for the pure Rashba or Dresselhaus spin-orbit system.  In Ref. [12], we calculated the transport properties of the special 2DEG with equal Rashba and Dresselhaus couplings, which possesses an equal-distant-like energy spectrum. The resonant spin Hall effect and accompanying resonant charge Hall effect also exhibit near the crossing between the nearest neighbor Landau levels. However, the most complex 2DEG with unequal Rashba and Dresselhaus couplings is not investigated up to now. In this work, we study the transport properties of this 2DEG, so that such semiconductor materials can be understood thoroughly.

The Hamiltonian for a single electron with spin-$\frac{1}{2}$ in a plane under a
perpendicular magnetic field is described by
$$ H=\frac{1}{2m^*}{\bf \Pi}^2-\frac{1}{2}g_s\mu_BB\sigma_z
+\frac{\alpha}{\hbar}(\sigma_y\Pi_x-\sigma_x\Pi_y)$$
 $$  +\frac{\beta}{\hbar}(\sigma_x\Pi_x-\sigma_y\Pi_y),\eqno{(1)}$$
where ${\bf \Pi}={\bf p}+\frac{e}{c}{\bf A}$, $\sigma_i(i=x, y, z)$ are the Pauli matrices
for electron spin, $g_s$ is the Lande g-factor,
$\mu_B$ is the Bohr magneton, $\alpha$ and $\beta$ represent the Rashba and the Dresselhaus spin-orbit couplings,
respectively. Here we have chosen the Landau guage ${\bf A}=yB\hat{{\bf x}}$. Because $[p_x, H]=0$, $p_x=k$ is a good quantum number.

The Hamiltonian (1) with a pure Rashba coupling (i.e. $\beta=0$) was solved by Rashba over fifty years ago [7]. The Landau levels of 2DEG with a pure Dresselhaus coupling (i.e. $\alpha=0$) can be obtained by the unitary transformation $\sigma_x\rightarrow\sigma_y$, $\sigma_y\rightarrow\sigma_x$ and $\sigma_z\rightarrow -\sigma_z$, which maps a 2DEG with Rashba coupling $\alpha$, Dresselhaus coupling $\beta$, and Lande g-factor $g_s$ to a 2DEG with Rashba coupling $\beta$, Dresselhaus coupling $\alpha$, and Lande g-factor $-g_s$ [13,14].
When the Rashba and Dresselhaus couplings coexist, the Hamiltonian (1) becomes  Extremely  complicated and its exact solution had been not obtained  for a long time. Fortunately, the eigenvalue problem was solved by using unitary transformations and introducing two bosonic annihilation operators $b_{k\sigma}=\frac{1}{\sqrt{2}l_c}[y+\frac{c}{eB}(k+ip_y)+\sqrt{2|a_R a_D|}u_\sigma]$ and the corresponding creation operators $b_{k\sigma}^\dagger=(b_{k\sigma})^\dagger$, with the cyclotron radius $l_c=\sqrt{\frac{\hbar c}{eB}}$, $a_R=\frac{\alpha m^*l_c}{\hbar^2}$, $a_D=\frac{\beta m^*l_c}{\hbar^2}$, $u_\sigma=\frac{\sqrt{2}}{2}\sigma[1-i {\rm sgn}(\alpha\beta)]$, and the orbital index $\sigma=\pm 1$ [11]. Different from the pure Rashba or Dresselhaus coupling case, the orbital space of electrons is divided into two independent infinite-dimensional subspaces described by the occupied number representations $\Gamma_\sigma$ associated with $b_{k\sigma}$ and $b^\dagger_{k\sigma}$. Then the Hamiltonian (1) can be rewritten as $H=H_{-1}\oplus H_1$, where $H_\sigma$ is the sub-Hamiltonian in $\Gamma_\sigma$. The eigenstates for the Hamiltonian (1) can be expressed as an infinite series in terms of the free Landau levels $\phi_{mk\sigma}$ (i.e. $b_{k\sigma}^\dagger \phi_{mk\sigma}=\sqrt{m+1}\phi_{m+1k\sigma}$, $b_{k\sigma} \phi_{mk\sigma}=\sqrt{m}\phi_{m-1k\sigma}$ and $<\phi_{m^\prime k\sigma^\prime}|\phi_{mk\sigma}>=\delta_{mm^\prime}\delta_{\sigma\sigma^\prime}$) in each $\Gamma_\sigma$ and the physical parameters.
Now the exact solution of the Hamiltonian (1) has been popularly accepted [14,15].

We note that the Rashba coupling $\alpha$ can be adjusted by a gate voltage perpendicular to the 2DEG plane [16,17]. Therefore, in experiments, an arbitrary ratio of two kinds of spin-orbit couplings can be obtained in different samples by changing the gate voltage. The relative strength of the Rashba and Dresselhaus couplings can be extracted from the photocurrent measurements [18,19]. Usually the coefficients $\alpha$ and $\beta$ have the same order of magnitude in quantum wells such as GaAs while in narrow gap compounds such as InAs, $\alpha$ dominates [16-19].

When $|\beta|=\rho|\alpha|$, the eigenvalue for the $n$th Landau level with the spin index $s$ and the orbital index $\sigma$ is given by $E_{ns}=\hbar\omega\epsilon_{ns}$ [11], where $\omega=\frac{eB}{m^*c}$ and
$$\epsilon_{ns}=n-\frac{\rho[2\rho a_R^2(1-\Delta_{ns}^2)-g\Delta_{ns}]}{1-\rho^2\Delta^2_{ns}}
+\frac{1}{2}(-1)^s$$
$$\times\sqrt{[1-\frac{4\rho a_R^2(1-\rho^2)\Delta_{ns}+g(1+\rho^2\Delta^2_{ns})}{1-\rho^2\Delta^2_{ns}}]^2+8na_R^2(1-\rho)^2}.\eqno{(2)}$$
Here the spin index $s=0, 1$ for $n\not=0$ while $s=0$ for $n=0$, $g=\frac{g_sm^*}{2m_e}$, and the energy level parameter $\Delta_{ns}$ is determined by the following equation
$$\{(1-\rho)[2(1+\rho)(1-\rho\Delta_{ns}^2)-\frac{g\Delta_{ns}}{a_R^2}][\epsilon_{ns}-n-\frac{1}{2}+a_R^2(1+\rho^2)]$$
$$+(1+\Delta_{ns})(1-\rho\Delta_{ns})(1-\rho^2-\frac{g}{2a_R^2})\}\sqrt{\rho}|a_R|=0.\eqno{(3)}$$
Obviously, when $\rho=0$, the eigenvalue (2) is nothing but that in 2DEG with pure Rashba spin-orbit interaction [7]. If $\rho=1$, the energy difference $E_{n0}-E_{n1}$ is independent of $n$, which means that the energy spectrum has an equal-distant-like structure.

The corresponding eigenstate is

$$|nks\sigma>=\frac{1}{{\cal N}_{ns\sigma}}\sum_{m=0}^{+\infty}
\left(
\begin{array}{cc}
1&\sqrt{\rho}\Delta_{ns}T^*_\sigma\\
-\sqrt{\rho}\Delta_{ns}T_\sigma&1
\end{array}\right)$$
$$\times\left (
\begin{array}{c}
\alpha_{ms}^{n\sigma}\\
T_\sigma\beta_{ms}^{n\sigma}
\end{array}\right )u^m_\sigma\phi_{mk\sigma},\eqno{(4)}$$
where ${\cal N}_{ns\sigma}$ is the normalized constant,
$|{\cal N}_{ns\sigma}|^2=(1+\rho\Delta^2_{ns})\sum_{m=0}^\infty(|\alpha_{ms}^{n\sigma}|^2
+|\beta_{ms}^{n\sigma}|^2),$
$T_\sigma=\frac{\sqrt{2}}{2}\sigma(sgn a_D+isgn a_R),$
and the components $\alpha_{ms}^{n\sigma}$ and $\beta_{ms}^{n\sigma}$ satisfy

$$\alpha_{ns}^{n\sigma}=\frac{\sqrt{2n} |a_R|{\cal C}_{ns} ({\cal C}_{ns}-{\cal B}_{ns})}
        {{\cal C}_{ns}{\cal D}_{ns}(\epsilon_{ns}-n-\lambda)+{\cal C}_{ns}\zeta_{ns}+\rho{\cal A}_{ns}\eta_{ns}}\beta_{n-1s}^{n\sigma},$$

$$\beta_{ns}^{n\sigma}=\frac{{\cal B}_{ns}{\cal D}_{ns}(\epsilon_{ns}-n-\lambda+1)+{\cal B}_{ns}\zeta_{ns}+\rho{\cal A}_{ns}\eta_{ns}}
       {\sqrt{2n}|a_R|{\cal B}_{ns}({\cal B}_{ns}-{\cal C}_{ns})}\alpha_{n-1s}^{n\sigma}\eqno{(5)}$$

at $m=n\not=0$, and

\begin{widetext}

$$\left (
\begin{array}{cc}
\sqrt{2m\rho} |a_R|{\cal A}_{ns}({\cal C}_{ns}-{\cal B}_{ns})&\sqrt{2m} |a_R|{\cal C}_{ns}({\cal C}_{ns}-{\cal B}_{ns})\\
-{\cal B}_{ns}{\cal D}_{ns}(\epsilon_{ns}-m-\lambda+1)-{\cal B}_{ns}\zeta_{ns}-\rho{\cal A}_{ns}\eta_{ns}&
\sqrt{\rho}[{\cal A}_{ns}{\cal D}_{ns}(\epsilon_{ns}-m-\lambda+1)-{\cal A}_{ns}\zeta_{ns}+{\cal B}_{ns}\eta_{ns}]
\end{array}
\right )
\left (
\begin{array}{c}
\alpha_{m-1s}^{n\sigma}\\
\beta_{m-1s}^{n\sigma}
\end{array}
\right )$$

$$=
\left (
\begin{array}{cc}
{\cal C}_{ns}{\cal D}_{ns}(\epsilon_{ns}-m-\lambda)+{\cal C}_{ns}\zeta_{ns}+\rho{\cal A}_{ns}\eta_{ns}&
-\sqrt{\rho}[{\cal A}_{ns}{\cal D}_{ns}(\epsilon_{ns}-m-\lambda)-{\cal A}_{ns}\zeta_{ns}+{\cal C}_{ns}\eta_{ns}]\\
\sqrt{2m\rho} |a_R| {\cal A}_{ns}({\cal C}_{ns}-{\cal B}_{ns})&\sqrt{2m} |a_R| {\cal B}_{ns}({\cal C}_{ns}-{\cal B}_{ns})
\end{array}
\right )
\left (
\begin{array}{c}
\alpha_{ms}^{n\sigma}\\
\beta_{ms}^{n\sigma}
\end{array}
\right ),\eqno{(6)}
$$
\end{widetext}
at $m\not=n$ and $m=n=0$. Here, we have defined
${\cal A}_{ns}=(1-\Delta_{ns})(1-\rho\Delta_{ns}),$
${\cal B}_{ns}=\rho(1-\Delta_{ns}^2),$
${\cal C}_{ns}=1-\rho^2\Delta_{ns}^2,$
${\cal D}_{ns}=1+\rho\Delta_{ns}^2,$
$\zeta_{ns}=\frac{1}{2}g(1-\rho\Delta_{ns}^2)+4\rho a_R^2(1+\rho)\Delta_{ns},$
$\eta_{ns}=2a_R^2(1+\rho)(1-\rho\Delta_{ns}^2)-g\Delta_{ns}$,
and $\lambda=2\rho a_R^2+\frac{1}{2}$.
Note that $\alpha_{-1s}^{n\sigma}=\beta_{-1s}^{n\sigma}=0$.
When $m \rightarrow + \infty,  \alpha_{ms}^{n\sigma}=\beta_{ms}^{n\sigma}=0$.

\begin{figure}
\rotatebox[origin=c]{0}{\includegraphics[angle=0,
           height=2.0in]{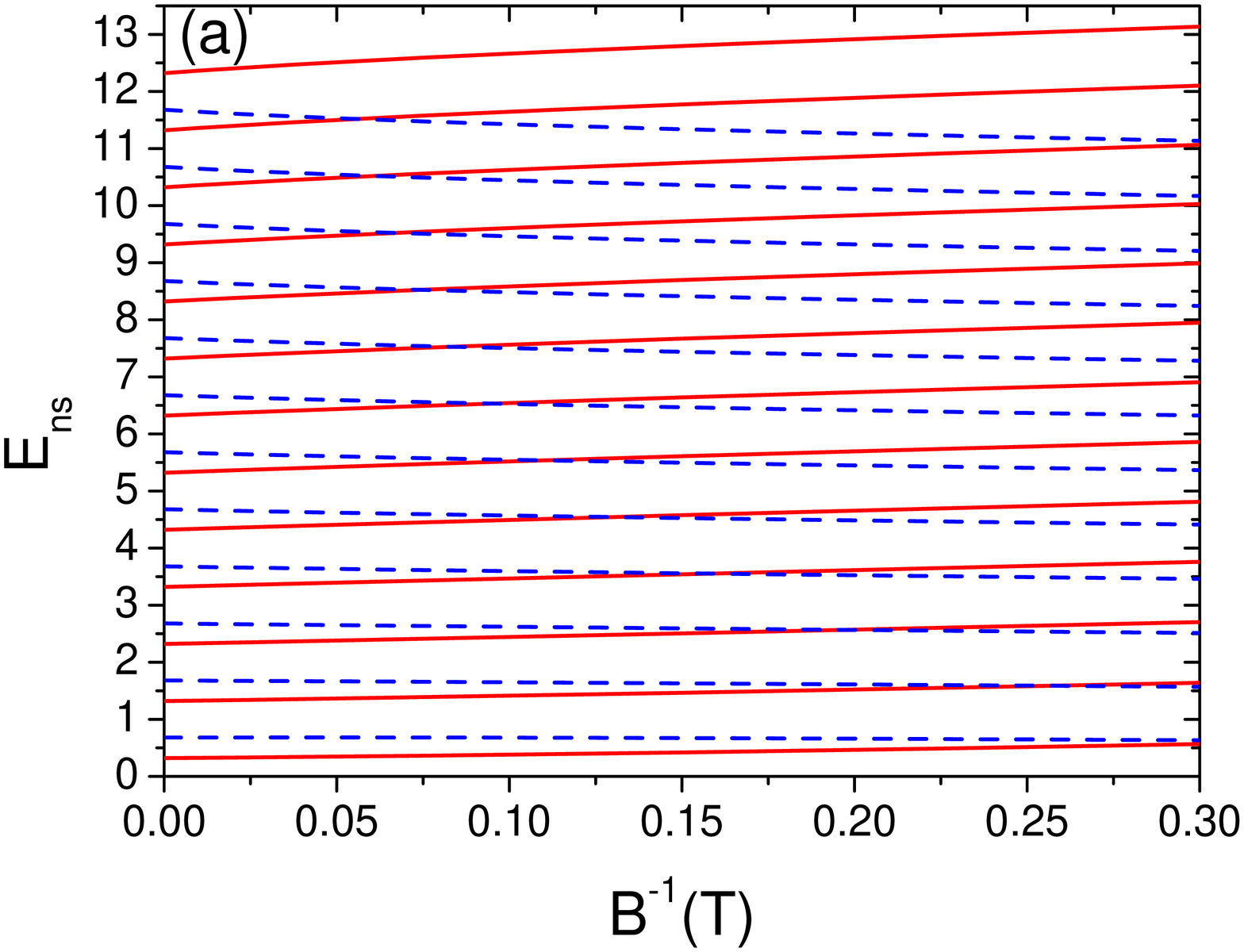}}
\rotatebox[origin=c]{0}{\includegraphics[angle=0,
           height=2.0in]{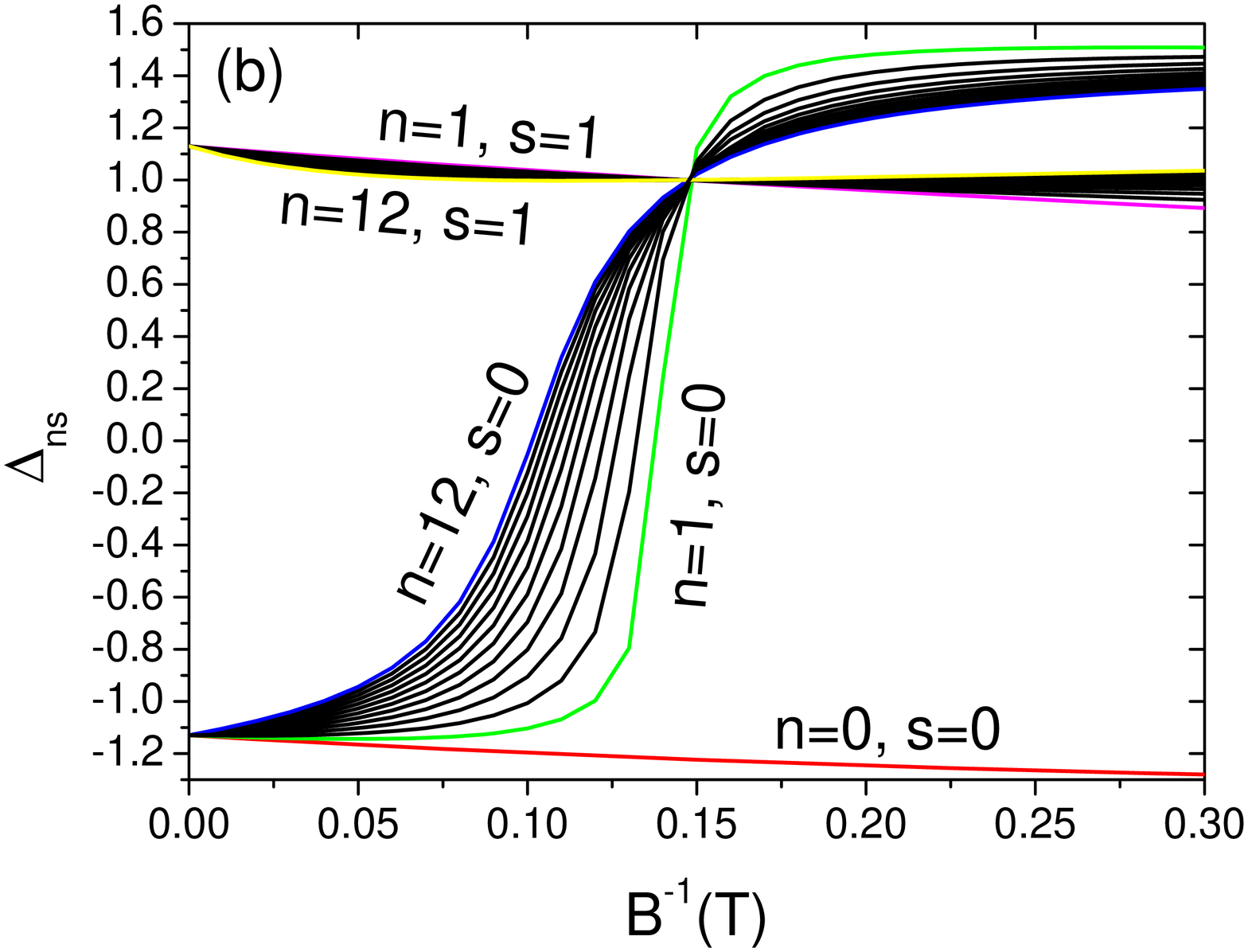}}
\caption {(Color online). (a) Landau energy levels $(n\leq 12)$ of an electron as a function of inverse magnetic field $1/B$ in the unit of $\hbar\omega$. The real line and dash line denote $s=0$ and $s=1$, respectively. (b) The corresponding energy level parameters. The material parameters used are $\rho=0.5$, $\alpha=4.0\times 10^{-11} eVm$, $m^*=0.05m_e$, and $g=0.1$. }
\end{figure}

Solving Eqs. (2) and (3), we can obtain the energy level parameter $\Delta_{ns}$ and the corresponding Landau level $E_{ns}$  at arbitrary magnetic field. In our calculations, we have chosen the typical material parameters $\rho=0.5$, $\alpha=4.0\times 10^{-11}$ eVm, $n_e=1.9\times 10^{16}/m^2$, $g_s=4$, and $m^*=0.05m_e$ [18,19]. In Fig. 1(a), we plot the Landau levels ($n\leq 12$) as a function of inverse magnetic field $1/B$.
Obviously, two nearest neighbor energy levels $E_{n0}$ and $E_{n+11}$ have a crossing at a certain magnetic field. The larger $n$, the stronger the
magnetic field at the crossing. Fig. 1(b) shows the energy level parameter $\Delta_{ns}$ associated with $E_{ns}$. We note that $\Delta_{n1} (n>0)$ and $\Delta_{00}$ vary slowly with increasing $1/B$ while $\Delta_{n0} (n>0)$ has a sharp change. When $1/B\rightarrow 0$, $\Delta_{n0}=-1.1298$ and $\Delta_{n1}=1.1298$. Interestingly, the curves of $\Delta_{ns}$ with $n>0$ meet at a point with a value of 0.9985 when $1/B \sim 0.1487 T^{-1}$.

Substituting $\Delta_{ns}$ and $E_{ns}$ into Eqs. (5) and (6), we get $\alpha_{ns}^{n\sigma}$ and $\beta_{ns}^{n\sigma}$, i.e. the eigenstates (4) for the complex 2DEG with $\rho=0.5$ in a wide range of magnetic fields. Then the expectation value of spin polarization per electron in this system is

$$S_z=\frac{\hbar}{4\nu}\sum_{nms\sigma}\frac{|a_R|f(E_{ns})}{|{\cal N}_{ns\sigma}|^2}\{(1-\rho\Delta^2_{ns})(|\alpha^{n\sigma}_{ms}|^2-
|\beta^{n\sigma}_{ms}|^2)$$
$$+2\sqrt{\rho}\Delta_{ns}[\alpha^{n\sigma}_{ms}(\beta^{n\sigma}_{ms})^*+
\beta^{n\sigma}_{ms}(\alpha^{n\sigma}_{ms})^*]\},\eqno{(7)}$$
where $\nu$ is the filling factor and $f(x)$ is the Fermi distribution function. The first term of the expression of $S_z$ is the contribution by both Rashba and Dresselhaus spin-orbit couplings while the second one is due to the coexistence of them, which plays a crucial role in the
magnetization of this 2DEG.
In Fig. 2(a), we present the curve of spin polarization $S_z$ with increasing the inverse magnetic field at zero temperature.
Because the Rashba and Dresselhaus spin-orbit interactions have different symmetries and compete with the Zeeman energy,
this 2DEG has a complex magnetization behavior, shown in Fig. 2(a).
We note that when $1/B\rightarrow 0$, $S_z\sim -0.28\hbar$. Therefore, such a system is hard to be magnetized.

In order to calculate the spin and charge transport properties in the 2DEG, we apply a small electric field $E$ along the $y$ axis. So an electron acquires an extra potential energy $H^\prime=eEy$, which can be treated as a perturbation term. The operator $y$ can be expressed in terms of the bosonic operators $b_\sigma$ and $b_\sigma^\dagger$ in each subspace $\Gamma_\sigma$, i.e. $y=\frac{\sqrt{2}}{2}l_c(b_{k\sigma}^\dagger+b_{k\sigma})-\frac{ck}{eB}-\sqrt{2\rho}|a_R|u_\sigma$. The charge current operator of a single electron in $\Gamma_\sigma$ reads $ j_{c\sigma}=-ev_{x\sigma}$
while the corresponding out of plane spin current operator in $\Gamma_\sigma$ is
$j_{s_z\sigma}=\frac{\hbar}{4}(\sigma_z v_{x\sigma}+v_{x\sigma}\sigma_z)$.
Here, the electron velocity in $\Gamma_\sigma$ along $x$ axis $v_{x\sigma}=\frac{1}{i\hbar}[x,H_\sigma+H^\prime_\sigma]=\frac{\hbar}
{\sqrt{2}m^*l_c}[b_{k\sigma}^\dagger+b_{k\sigma}+\sqrt{2}|a_R|({\rm sgn}\alpha\sigma_y +\rho{\rm sgn}\beta \sigma_x -\frac{\sqrt{\rho}\sigma}{l_c})]$.
Up to the first order in the electric field $E$ in the expectation value of the spin or charge current operator,
the spin or charge Hall conductance, i.e. the coefficient of the linear term,  can be expressed as [10]

$$G_{s_z,c}=\frac{1}{E}\sum_{nn^{\prime}kss^{\prime}\sigma}\frac{(H^\prime_\sigma)_{nks\sigma}^{n^\prime ks^\prime\sigma}(j_{s_z\sigma,c\sigma})_{n^\prime ks^\prime\sigma}^{nks\sigma}}
  {E_{ns}-E_{n^\prime s^\prime}}f(E_{ns})+{\rm h.c.},
\eqno{(8)}$$
where the matrix elements $(H^\prime_\sigma)_{nks\sigma}^{n^\prime ks^\prime\sigma}=<n^\prime ks^\prime\sigma|H^\prime_\sigma|nks\sigma>$, $(j_{s_z\sigma})^{n^\prime ks^\prime\sigma}_{nks\sigma}=<n^\prime ks^\prime\sigma|j_{s_z\sigma}|nks\sigma>$, and $(j_{c\sigma})^{n^\prime ks^\prime\sigma}_{nks\sigma}=<n^\prime ks^\prime\sigma|j_{c\sigma}|nks\sigma>$ can be obtained by using the eigenvalue (2), the energy level parameter (3), and the eigenstate (4). Obviously, $G_{s_z}$ ang $G_{c}$ are highly nonlinear functions in terms of the material parameters $\rho$, $|a_R|$, $g$, and the magnetic field $B$, which reveal complex transport characteristics in this system.

\begin{figure}
\rotatebox[origin=c]{0}{\includegraphics[angle=0,
           height=2in]{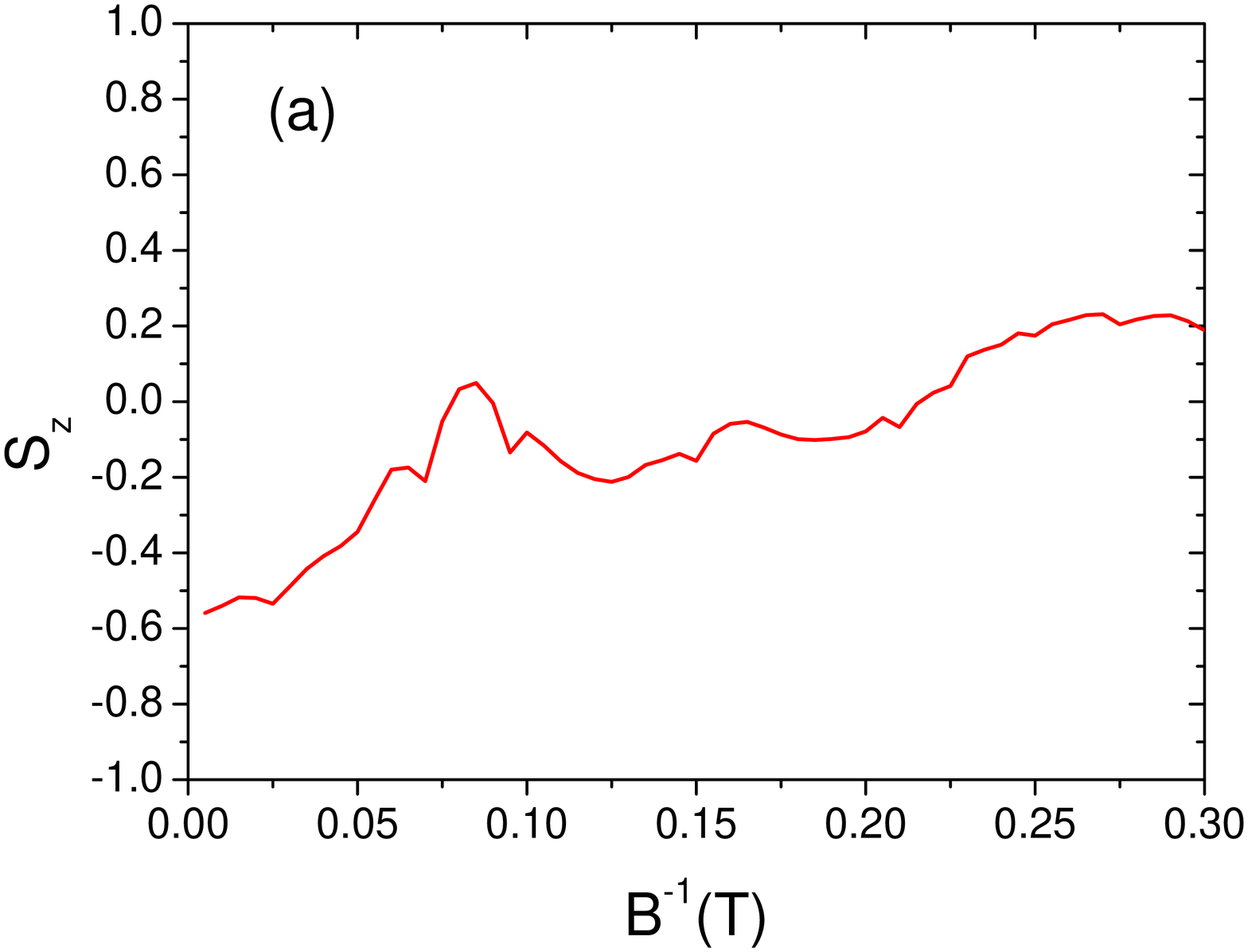}}
\rotatebox[origin=c]{0}{\includegraphics[angle=0,
           height=2in]{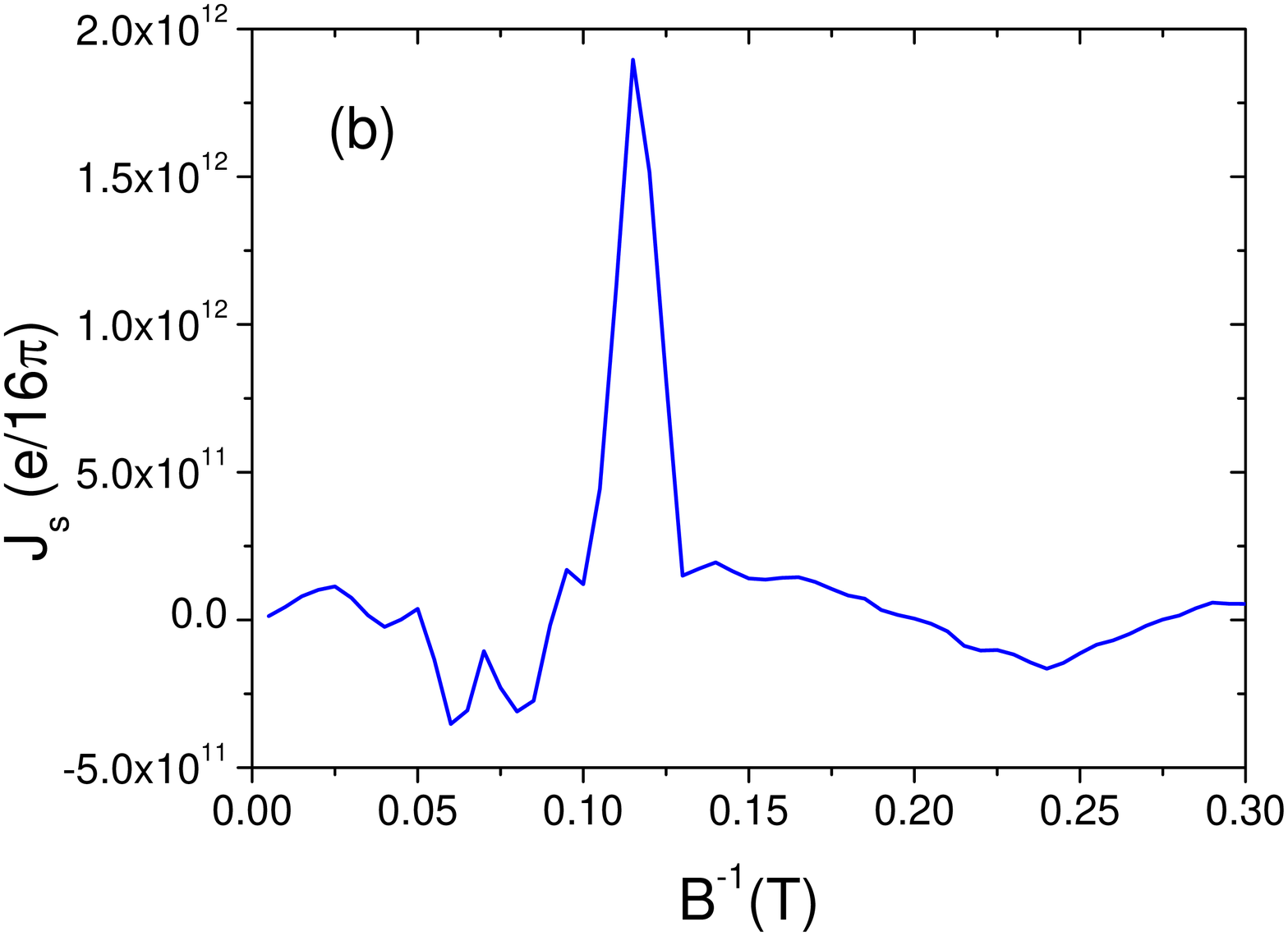}}
\rotatebox[origin=c]{0}{\includegraphics[angle=0,
           height=2in]{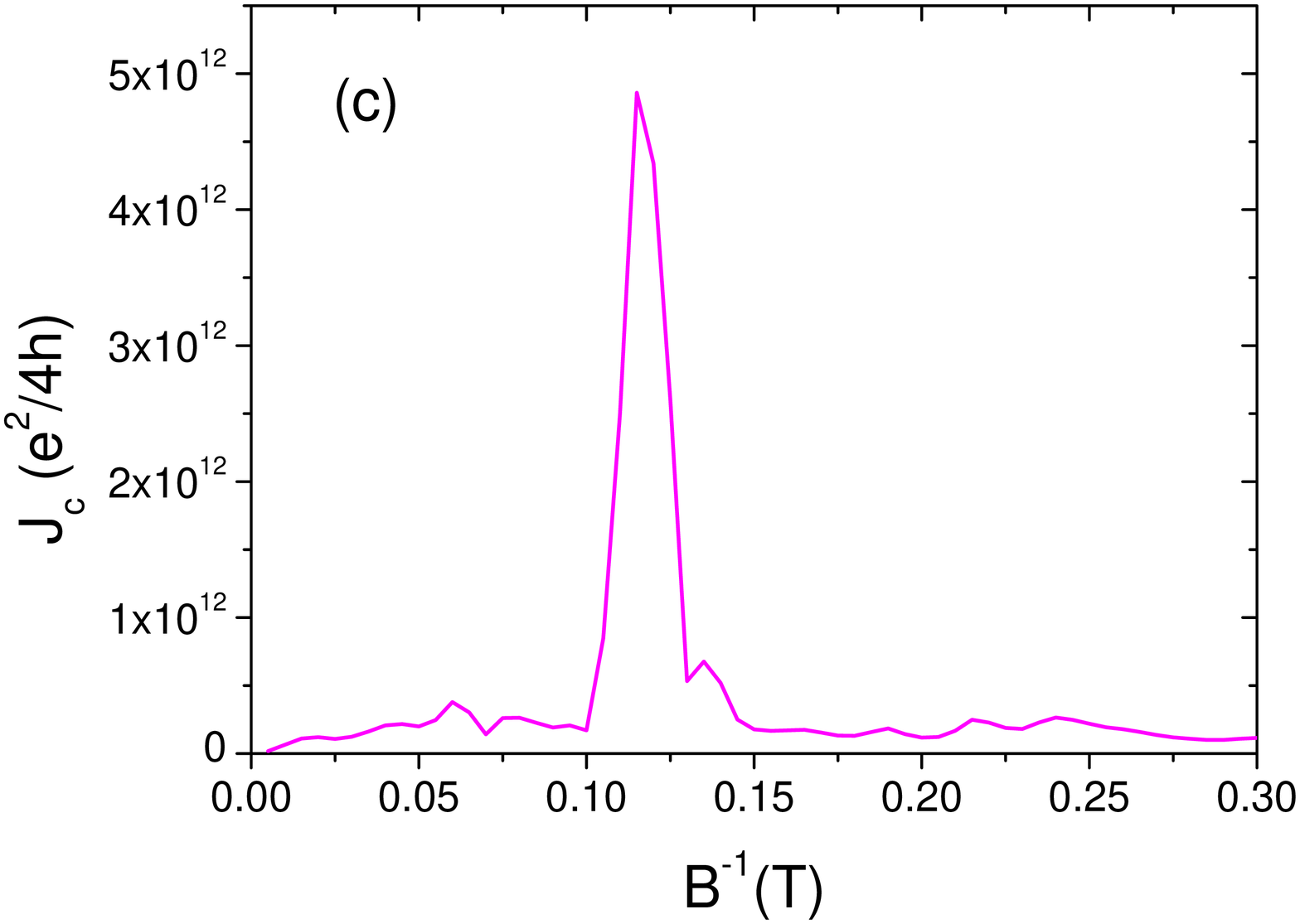}}
\caption{$S_Z$ (unit: $\hbar/2$), $G_{s_z}$ and $G_c$ as a function of $1/B$ at zero temperature. Here the electron density $n_e=1.9\times 10^{16} m^{-2}$ and the other physical parameters are same with those in Fig. 1.}
\end{figure}

After a tedious but straight calculation we depict the curves of $G_{s_z}$ and $G_{c}$ as a function of $1/B$ at zero temperature in Figs. 2(b) and 2(c), respectively. It is clear that $G_{s_z}$ and $G_{c}$  possess simultaneously a resonant peak at near $B_c\sim (1/0.13) T$, at which the Landau levels $E_{40}$ and $E_{51}$ cross. When $B\rightarrow (1/0.115) T$, the electrons begin to fill the energy level $E_{51}$, and $G_{s_z}$ and $G_{c}$
sharply increase. However, at $B_c$, $E_{40}$ and $E_{51}$ are filled fully while $E_{61}$ is filled partly, and $G_{s_z}$ and $G_{c}$ decrease rapidly. Therefore, the resonant spin and charge Hall conductances are produced by filling partly two nearest neighbor Landau levels near their crossing point. Such resonant phenomena also happen in the 2DEG with pure Rashba or equal Rashba and Dresselhaus spin-orbit coupling [10,12].

In summary, we have calculated the spin polarization, spin and charge Hall conductances in 2DEG in the presence of unequal Rashba and Dresselhaus spin-orbit interactions under a perpendicular magnetic field by employing the exact solution for the Hamiltonian (1). The coexistence of the
resonant spin and charge Hall effects appears at the vicinity of the crossing point of two nearest neighbor Landau levels, which are not filled fully by electrons. We also note that the spin polarization and both Hall conductances are determined only by the magnitudes of the spin-orbit couplings $\alpha$ and $\beta$ and are independent of their signs. It is expected that these resonant phenomena could have potential application for semiconductor spintronics.

This work was supported by the Sichuan Normal University, the "Thousand Talented Program" of Sichuan Province, China, and the Texas Center for Superconductivity at the University of Houston and by the Robert A. Welch Foundation under grant No. E-1146.


\begin{thebibliography}{99}

\bibitem{1} S. Murakami, N. Nagaosa, and S. C. Zhang, Science {\bf 301}, 1348 (2003).
\bibitem{2} J. Sinova, D. Culcer, Q. Niu, N. A. Sinitsyn, T. Jungwirth, and A. H. MacDonald, Phys. Rev. Lett. {\bf 92}, 126603 (2004).
\bibitem{3} Y. K. Kato, R. C. Myers, A. C. Gossard, and D. D. Awschalom, Science {\bf 306}, 1910 (2004).
\bibitem{4} J. Wunderlich, B. Kaestner, J. Sinova, and T. Jungwirth, Phys. Rev. Lett. {\bf 94}, 047204 (2005).
\bibitem{5} V. Sih, R. C. Myers, Y. K. Kato, W. H. Lau, A. C. Gossard and D. D. Awschalom, Nature Physics, {\bf 1}, 31 (2005).
\bibitem{6} S. O. Valenzuela and M. Tinkham, Nature {\bf 442}, 176 (2006).
\bibitem{7} E. I. Rashba, Sov. Phys. Solid State {\bf 2}, 1109 (1960).
\bibitem{8} G. Dresselhaus, Phys. Rev. B {\bf 100}, 580 (1955).
\bibitem{9} S. Datta and B. Das, Appl. Phys. Lett. {\bf 56}, 665 (1990).
\bibitem{10} Shun-Qing Shen, Michael Ma, X. C. Xie, and Fu-Chun Zhang, Phys. Rev. Lett. {\bf 92}, 256603 (2004).
\bibitem{11} Degang Zhang, J. Phys. A: Math. Gen. {\bf 39}, L477 (2006).
\bibitem{12} Degang Zhang, Yao-Ming Mu, and C. S. Ting, Appl. Phys. Lett. {\bf 92}, 212103 (2008).
\bibitem{13} Shun-Qing Shen, Yun-Juan Bao, Michael Ma, X. C. Xie, and Fu-Chun Zhang, Phys. Rev. B {\bf 71}, 155316 (2005).
\bibitem{14} Fu-Chun Zhang and Shun-Qing Shen, Inter. J. Mod. Phys. B {\bf 22}, 94 (2008).
\bibitem{15} B. Estienne, S. M. Haaker, and K. Schoutens, New J. Phys. {\bf 13}, 045012 (2011).
\bibitem{16} J. Nitta, T. Akazaki, H. Takayanagi, and T. Enoki, Phys. Rev. Lett. {\bf 78}, 1335 (1997).
\bibitem{17} J. B. Miller, D. M. Zumbuhl, C. M. Marcus, Y. B. Lyanda-Geller, D. Goldhaber-Gordon, K. Campman, and A. C. Gossard, Phys. Rev. Lett. {\bf 90}, 076807 (2003).
\bibitem{18} S. D. Ganichev, V. V. Bel'kov, L. E. Golub, E. L. Ivchenko, P. Schneider, S. Giglberger, J. Eroms, J. De Boeck, G. Borghs, W. Wegscheider, D. Weiss, and W. Prettl, Phys. Rev. Lett. {\bf 92}, 256601 (2004).
\bibitem{19} S. Giglberger, L. E. Golub, V. V. Bel'kov, S. N. Danilov, D. Schuh, Ch. Gerl, F. Rohlfing, J. Stahl,
W. Wegscheider, D. Weiss, W. Prettl, and S. D. Ganichev, Phys. Rev. B {\bf 75}, 035327 (2007).


\end{thebibliography}
\end{document}